\documentstyle[12pt,axodraw]{article}

\newcommand{\mysection}{\setcounter{equation}{0}\section}

\begin{document}
\vskip 0.2cm
\hfill{YITP-SB-04-46}
\vskip 0.2cm
\centerline{\large\bf {Two-loop corrections to Higgs boson production}} 
\vskip 0.4cm
\centerline {\sc V. Ravindran}
\centerline{\it Harish-Chandra Research Institute,}
\centerline{\it Chhatnag Road, Jhusi,}
\centerline{\it Allahabad, 211019, India.}
\vskip 0.2cm
\centerline {\sc J. Smith 
\footnote{partially supported
by the National Science Foundation grant PHY-0098527.}
}
\centerline{\it C.N. Yang Institute for Theoretical Physics,}
\centerline{\it State University of New York at Stony Brook,
New York 11794-3840, USA.}
\vskip 0.2cm
\centerline {\sc W.L. van Neerven}
\centerline{\it Instituut-Lorentz}
\centerline{\it University of Leiden,}
\centerline{\it PO Box 9506, 2300 RA Leiden,}
\centerline{\it The Netherlands.}
\vskip 0.2cm
\centerline{August 2004}
\vskip 0.2cm
\centerline{\bf Abstract}
\vskip 0.3cm
In this paper we present the complete two-loop vertex corrections to scalar
and pseudo-scalar Higgs
boson production for general colour factors for the gauge group ${\rm SU(N)}$
in the limit where the top quark mass gets infinite. 
We derive a general formula for the vertex correction which holds
for conserved and non conserved operators. For the conserved operator
we take the electromagnetic vertex correction as an example whereas
for the non conserved operators we take the two vertex corrections above.
Our observations for the structure of the pole terms $1/\varepsilon^4$,
$1/\varepsilon^3$ and $1/\varepsilon^2$ in two loop order are the same as 
made earlier in the literature for electromagnetism.
However we also elucidate the origin of the second order single pole term
which is equal to the second order singular part of the anomalous dimension
plus a universal function which is the same for the quark and the gluon.\\[3mm]


\vfill

\mysection{Introduction}
One of the crucial tests of the standard model will be the discovery of the
Higgs boson at the LHC or at the TEVATRON. Its discovery or its
absence will shed light on the mechanism how particles acquire their
mass. It will also answer questions about supersymmetric extensions of the
standard model or about compositeness of the existing particles including
the Higgs boson. In this paper we will concentrate on Higgs production
where the lowest order reaction proceeds via the gluon-gluon fusion
mechanism. In the standard model the scalar Higgs boson H couples to the gluons
via heavy quarks among which the top-quark is the most prominent one.
This also holds for the pseudo-scalar Higgs boson A provided $\tan \beta$ is 
small where $\beta$ denotes the mixing angle in the Two-Higgs-Doublet model
(2HDM). The basic production mechanism is that two incoming gluons are 
coupled to the Higgs boson through a triangular top-quark loop \cite{wil}.
The first order QCD corrections to this process \cite{gsz} are already very 
complicated
in particular the virtual corrections which involve massive two-loop 
integrals. However fortunately one discovered that one could make a 
simplification for the total cross section \cite{gsz}, \cite{dawson}. 
The latter is also possible
for differential distributions provided the transverse momentum $p_T$ is not
too large ($p_T\le 2\,m_H$) \cite{ehsb}, \cite{fgk}, \cite{rasm1}, \cite{glsc}. This is achieved by 
taking the large top-quark
mass limit $m_t\rightarrow \infty$. In this case the Feynman graphs are derived
from an effective Lagrangian describing the direct coupling of the (pseudo-)
scalar Higgs boson to the gluons. Therefore the triangular top-quark loop
is replaced by a simple vertex and the treatment of the total cross section
is the same as for the Drell-Yan process. In this paper we concentrate 
on the one and two-loop vertex QCD corrections to the total cross section for
Higgs boson production. For the two-loop electroweak corrections we refer
to \cite{agl} and \cite{deg}.
Although the two-loop virtual contributions for scalar Higgs production
in the heavy top-quark limit were presented in \cite{harland},
the result was not decomposed in the various colour factors .
This is important for the general structure of the vertex function as
will be pointed out later on. An explicit expression for the 
two-loop virtual corrections to the pseudo-scalar Higgs boson is
not given anywhere although it is indirectly contained in \cite{haki3}, 
\cite{anme2}. In this paper we explicitly present the two-loop virtual 
contributions to scalar as well as pseudo-scalar Higgs boson production
for general colour factors in ${\rm SU(N)}$. 

The paper will be organized as follows.
In section 2 we give an outline of the renormalization constants which are
involved in the calculation of the two-loop virtual corrections to
scalar and pseudo-scalar Higgs production. In section 3 we present the 
calculation and compute the contribution to the flavour dependent part
of the coefficient function in the total cross section for (pseudo-) scalar
Higgs production. In section 4 we discuss the structure of the vertex 
correction w.r.t. the pole terms and present the finite terms. In particular
we find that also the single pole term can be predicted so that the
vertex and other two-loop radiative corrections can be expressed in a unique form.
For completeness we also compare it with the electromagnetic
form factor present e.g. in the Drell-Yan process.

\mysection{Application of the effective Lagrangian approach to Higgs 
production.}
In the large top-quark mass limit the Feynman rules (see e.g. \cite{kdr})
for scalar Higgs  production (${\rm H}$) can be derived from the following
effective Lagrangian density
\begin{eqnarray}
\label{eqn2.1}
{\cal L}^{\rm H}_{eff}=G_{\rm H}\,\Phi^{\rm H}(x)\,O(x) \quad
\mbox{with} \quad O(x)=-\frac{1}{4}\,G_{\mu\nu}^a(x)\,G^{a,\mu\nu}(x)\,,
\end{eqnarray}
whereas pseudo-scalar Higgs (${\rm A}$) production is obtained from
\begin{eqnarray}
\label{eqn2.2}
&&{\cal L}_{eff}^{\rm A}=\Phi^{\rm A}(x)\Bigg [G_{\rm A}\,O_1(x)+
\tilde G_{\rm A}\,O_2(x)\Bigg ] \quad \mbox{with} \quad
\nonumber\\[2ex]
&&O_1(x)=-\frac{1}{8}\,\epsilon_{\mu\nu\lambda\sigma}\,G_a^{\mu\nu}\,
G_a^{\lambda\sigma}(x) \,,
\nonumber\\[2ex]
&&O_2(x) =-\frac{1}{2}\,\partial^{\mu}\,\sum_{i=1}^{n_f}
\bar q_i(x)\,\gamma_{\mu}\,\gamma_5\,q_i(x)\,,
\end{eqnarray}
where $\Phi^{\rm H}(x)$ and  $\Phi^{\rm A}(x)$ represent the scalar and
pseudo-scalar fields respectively and $n_f$ denotes the number of light
flavours.
Furthermore the gluon field strength is given by $G_a^{\mu\nu}$ and the
quark field is denoted by $q_i$.
The factors multiplying the operators are chosen in such a way that the
vertices are normalised to the effective coupling constants $G_{\rm H}$,
$G_{\rm A}$ and $\tilde G_{\rm A}$. The latter are determined by the
top-quark triangular loop graph, including all QCD corrections, taken in the
limit $m_t\rightarrow \infty$ which describes the decay process
${\rm B} \rightarrow g + g$ with ${\rm B}={\rm H},{\rm A}$ namely
\begin{eqnarray}
\label{eqn2.3}
G_{\rm B}&=&-2^{5/4}\,a_s(\mu_r^2)\,G_F^{1/2}\,
\tau_{\rm B}\,F_{\rm B}(\tau_{\rm B})\,{\cal C}_{\rm B}
\left (a_s(\mu_r^2),\frac{\mu_r^2}{m_t^2}\right )\,,
\nonumber\\[2ex]
\tilde G_{\rm A}&=&-\Bigg [a_s(\mu_r^2)\,C_F\,\left (\frac{3}{2}-3\,
\ln \frac{\mu_r^2}{m_t^2}\right )+\cdots \Bigg ]\,G_{\rm A}\,,
\end{eqnarray}
and $a_s(\mu_r^2)$ is the renormalized coupling constant defined by
\begin{eqnarray}
\label{eqn2.4}
a_s(\mu_r^2)=\frac{\alpha_s(\mu_r^2)}{4\pi}\,,
\end{eqnarray}
where $\alpha_s(\mu_r^2)$ is the running coupling constant and $\mu_r$ denotes
the renormalization scale. Further $G_F$ represents the Fermi constant and the
functions $F_{\rm B}$ are given by
\begin{eqnarray}
\label{eqn2.5}
&& F_{\rm H}(\tau)=1+(1-\tau)\,f(\tau)\,, \qquad  F_{\rm A}(\tau)=f(\tau)\,\cot
\beta\,,
\nonumber\\[2ex]
&& \tau=\frac{4\,m_t^2}{m^2} \,,
\nonumber\\[2ex]
&&f(\tau)=\arcsin^2 \frac{1}{\sqrt\tau}\,, \quad \mbox{for} \quad \tau \ge 1\,,
\nonumber\\[2ex]
&& f(\tau)=-\frac{1}{4}\left ( \ln \frac{1-\sqrt{1-\tau}}{1+\sqrt{1-\tau}}
+\pi\,i\right )^2 \quad \mbox{for} \quad \tau < 1\,,
\end{eqnarray}
where $\cot \beta$ denotes the mixing angle in the Two-Higgs-Doublet Model.
Further $m$ and $m_t$ denote the masses of the (pseudo-) scalar Higgs
boson and the top quark respectively. In the large $m_t$-limit
we have
\begin{eqnarray}
\label{eqn2.6}
 \mathop{\mbox{lim}}\limits_{\vphantom{\frac{A}{A}} \tau \rightarrow \infty}
F_{\rm H}(\tau)=\frac{2}{3\,\tau}\,, \qquad
 \mathop{\mbox{lim}}\limits_{\vphantom{\frac{A}{A}} \tau \rightarrow \infty}
F_{\rm A}(\tau)=\frac{1}{\tau}\,\cot \beta\,.
\end{eqnarray}
The coefficient functions ${\cal C}_{\rm B}$ originate from the corrections
to the top-quark triangular graph provided one takes the
limit $m_t\rightarrow \infty$.  We have presented the Born level
couplings $G_{\rm B}$ in Eq. (\ref{eqn2.3}) for general $m_t$ for on-shell
gluons only in order to keep some part of the top-quark mass
dependence. This is an approximation because the gluons which couple to the
H and A bosons via the top-quark loop in partonic
subprocesses are very often virtual. The virtual-gluon momentum dependence is
neither described by $F_{\rm B}(\tau)$ nor by ${\cal C}_{\rm B}$. However
for total cross sections the main contribution comes from the region
where the gluons are almost on-shell so that this approximation is
better than it is for differential cross sections with
large transverse momentum. The coefficient functions are computed
up to order $\alpha_s^2$ in \cite{kls}, \cite{cks} for the H
and in \cite{cksb} for the A. They read as follows
\begin{eqnarray}
\label{eqn2.7}
&&{\cal C}_{\rm H}\left (a_s(\mu_r^2),\frac{\mu_r^2}{m_t^2}\right )=
1+a_s^{(5)}(\mu_r^2)\,\Bigg [5\,C_A-3\,C_F\Bigg ] +
\left (a_s^{(5)}(\mu_r^2)\right )^2\,\Bigg [
\frac{27}{2}\,C_F^2
\nonumber\\[2ex]
&&-\frac{100}{3}\,C_A\,C_F+\frac{1063}{36}\,C_A^2-\frac{4}{3}\,
C_F\,T_f-\frac{5}{6}\,C_A\,T_f +\Big (7\,C_A^2
\nonumber\\[2ex]
&& -11\,C_A\,C_F\Big )\,\ln \frac{\mu_r^2}{m_t^2}
+ n_f\,T_f\,\left (-5\,C_F-\frac{47}{9}\,C_A
+8\,C_F\,\ln \frac{\mu_r^2}{m_t^2} \right ) \Bigg ]\,,
\\[2ex]
\label{eqn2.8}
&&{\cal C}_{\rm A}\left (a_s(\mu_r^2),\frac{\mu_r^2}{m_t^2}\right )=1\,,
\end{eqnarray}
where $a_s^{(5)}$ is presented in the five-flavour number scheme. Notice that
the coefficient function in Eq. (2.7) is derived for general colour factors
of the group SU(N) from Eq. (6) in \cite{cks}. These factors are given by
\begin{eqnarray}
\label{eqn2.9}
C_A=N\quad\,,\quad C_F=\frac{N^2-1}{2N}\quad\,,\quad T_f=\frac{1}{2}\,.
\end{eqnarray}
Notice that $T_f$ is also incorporated into $G_{\rm B}$ in Eq. (\ref{eqn2.3})
where it is set to the value $T_f=1/2$.
Using the effective Lagrangian approach we will calculate the
total cross section of the reaction
\begin{eqnarray}
\label{eqn2.10}
H_1(P_1)+H_2(P_2)\rightarrow {\rm B}(q)+'X'\,,
\end{eqnarray}
where $H_1$ and $H_2$ denote the incoming hadrons and $X$ represents an
inclusive hadronic state. The total cross section is given by
\begin{eqnarray}
\label{eqn2.11}
&&\sigma_{\rm tot}=\frac{\pi\,G_{\rm B}^2}{8\,(N^2-1)}\,\sum_{a,b=q,\bar q,g}\,
\int_x^1 dx_1\, \int_{x/x_1}^1dx_2\,f_a(x_1,\mu^2)\,f_b(x_2,\mu^2)\,
\nonumber\\[2ex] && \qquad\qquad\times
\Delta_{ab,{\rm B}}\left ( \frac{x}{x_1\,x_2},\frac{m^2}{\mu^2} \right ) \,,
\quad {\rm B}={\rm H},{\rm A}\,,
\nonumber\\[2ex]
&&\mbox{with}\quad x=\frac{m^2}{S} \quad\,,\quad S=(P_1+P_2)^2\quad
\,,\quad q^2=m^2\,,
\end{eqnarray}
where the factor $1/(N^2-1)$ originates from the colour average
in the case of the local gauge group SU(N). Further
we have assumed that the (pseudo-) scalar Higgs boson is mainly
produced on-shell i.e. $q^2\sim m^2$. The parton densities denoted by
$f_a(y,\mu^2)$ ($a,b=q,\bar q,g$)
depend on the mass factorization/renormalization scale $\mu$.
The same scale also enters the coefficient functions $\Delta_{ab,{\rm B}}$
which are derived from the partonic cross sections ($z=m^2/s$)
\begin{eqnarray}
\label{eqn2.12}
&&\sigma_{ab,{\rm B}}\left (z,\frac{m^2}{\mu^2}\right )=\frac{\pi}{s}\,
\int \frac{d^nq}{(2\pi)^n}\,\delta^+(q^2-m^2)\,
T_{ab,{\rm B}}(q,p_1,p_2)\,,
\end{eqnarray}
where the incoming partons $a$ and $b$ carry momenta $p_1$ and $p_2$ 
respectively. They are related to the hadron momenta by
\begin{eqnarray}
\label{eqn2.13}
&&p_1=x_1\,P_1\quad\,,\quad p_2=x_2\,P_2 \quad\,,\quad
\nonumber\\[2ex]
&&s=(p_1+p_2)^2 \quad,\quad \Longrightarrow \qquad  s=x_1\,x_2\,S\,,
\qquad z=\frac{m^2}{s}\,.
\end{eqnarray}
The amplitude $T_{ab,{\rm B}}$ can be written as
\begin{eqnarray}
\label{eqn2.14}
T_{ab,{\rm H}}(q,p_1,p_2)&=&G_{\rm H}^2\int d^4y\,e^{i\,q\cdot y}
\, \langle a,b|O(y)\,O(0)|a,b\rangle \,,
\\[2ex]
\label{eqn2.15}
T_{ab,{\rm A}}(q,p_1,p_2)&=&\int d^4y\,e^{i\,q\cdot y}\,
\langle a,b|\Big (G_{\rm A}\,O_1(y)+\tilde G_{\rm A}\,O_2(y)\Big )
\nonumber\\[2ex]
&&\times\Big (G_{\rm A}\,O_1(0)+\tilde G_{\rm A}\,O_2(0)\Big )
|a,b\rangle \,.
\end{eqnarray}
The expressions above for the amplitude $T_{ab}$
are similar to those given for the Drell-Yan process except that the
conserved electroweak currents are replaced by the operators $O$ and $O_1,O_2$.
The latter are not conserved so that they acquire additional renormalization
constants defined by
\begin{eqnarray}
\label{eqn2.16}
O(y)=Z_O\,\hat O(y) \quad\,,\quad O_i(y)=Z_{ij}\, \hat O_j(y)\,.
\end{eqnarray}
where the hat indicates that the quantities under consideration are
unrenormalized.
Insertion of the above equations into Eqs. (\ref{eqn2.14}),(\ref{eqn2.15})
leads to the renormalized expressions
\begin{eqnarray}
\label{eqn2.17}
&&T_{ab,{\rm H}}(q,p_1,p_2)=\,G_{\rm H}^2\,Z_O^2\,\int d^4y\,
e^{i\,q\cdot y} \, \langle a,b|\hat O(y)\,\hat O(0)|a,b\rangle\,,
\\[2ex]
\label{eqn2.18}
&&T_{ab,{\rm A}}(q,p_1,p_2)=\int d^4y\,e^{i\,q\cdot y}\,
\Bigg [\Bigg \{G_{\rm A}^2\,Z_{11}^2+\tilde G_{\rm A}^2\,Z_{21}^2+
2\,G_{\rm A}\,\tilde G_{\rm A}\,Z_{11}\,Z_{21}\Bigg \}
\nonumber\\[2ex]
&&\times \langle a,b|\hat O_1(y)\,\hat O_1(0)|a,b\rangle+\Bigg \{ G_{\rm A}^2
\,Z_{12}^2 +\tilde G_{\rm A}^2\,Z_{22}^2
+2\,G_{\rm A}\,\tilde G_{\rm A}\,Z_{12}\,Z_{22}\Bigg \}
\nonumber\\[2ex]
&&\times \langle a,b|\hat O_2(y)\,\hat O_2(0)|a,b\rangle
+\Bigg \{ G_{\rm A}^2\,Z_{11}\,Z_{12}
+\tilde G_{\rm A}^2\,Z_{22}\,Z_{21}+G_{\rm A}\,\tilde G_{\rm A}
\nonumber\\[2ex]
&&\times \Big (Z_{11}\,Z_{22}+Z_{12}\,Z_{21}\Big )\Bigg \}\,
\langle a,b|\hat O_1(y)\,\hat O_2(0) +\hat O_2(y)\,\hat O_1(0)|a,b\rangle
\Bigg ]\,.
\end{eqnarray}
Since we are interested in the one- and two-loop correction to the
subprocess $B\rightarrow g + g$ the above formula simplifies and it 
becomes
\begin{eqnarray}
\label{eqn2.19}
T_{gg,{\rm H}}(q,p_1,p_2)&=&\,G_{\rm H}^2\,Z_O^2\,\int d^4y\,
e^{i\,q\cdot y} \, \langle g,g|\hat O(y)\,\hat O(0)|g,g\rangle\,,
\\[2ex]
\label{eqn2.20}
T_{ab,{\rm A}}(q,p_1,p_2)&=&\int d^4y\,e^{i\,q\cdot y}\,
\Bigg [G_{\rm A}^2\,Z_{11}^2\,\langle g,g|\hat O_1(y)\,\hat O_1(0)|g,g\rangle
+\Bigg \{ G_{\rm A}^2\,Z_{12}
\nonumber\\[2ex]
&&+G_{\rm A}\,\tilde G_{\rm A}\Bigg \}\,
\langle g,g|\hat O_1(y)\,\hat O_2(0) +\hat O_2(y)\,\hat O_1(0)|g,g\rangle
\Bigg ]\,.
\end{eqnarray}
The operator renormalization constants depend on the regularization scheme
in particular on the
prescription for the $\gamma_5$-matrix and the Levi-Civita tensor in
Eq. (\ref{eqn2.2}). The computation of $T_{gg,{\rm H}}$ will be carried
out by choosing n-dimensional regularization where in the case of
$T_{gg,{\rm A}}$ we adopt the HVBM prescription
\cite{hove}, \cite{brma} for the $\gamma_5$-matrix. For this choice
the contraction of the Levi-Civita tensors proceeds as
\begin{eqnarray}
\label{eqn2.21}
\epsilon_{\mu_1\nu_1\lambda_1\sigma_1}\,\epsilon^{\mu_2\nu_2\lambda_2\sigma_2}=
\large{\left |
\begin{array}{cccc}
\delta_{\mu_1}^{\mu_2}&\delta_{\mu_1}^{\nu_2}&\delta_{\mu_1}^{\lambda_2}&
\delta_{\mu_1}^{\sigma_2}\\
\delta_{\nu_1}^{\mu_2}&\delta_{\nu_1}^{\nu_2}&\delta_{\nu_1}^{\lambda_2}&
\delta_{\nu_1}^{\sigma_2}\\
\delta_{\lambda_1}^{\mu_2}&\delta_{\lambda_1}^{\nu_2}&
\delta_{\lambda_1}^{\lambda_2}&\delta_{\lambda_1}^{\sigma_2}\\
\delta_{\sigma_1}^{\mu_2}&\delta_{\sigma_1}^{\nu_2}&
\delta_{\sigma_1}^{\lambda_2}&\delta_{\sigma_1}^{\sigma_2}
\end{array}
\right |}\,,
\end{eqnarray}
where all Lorentz indices are taken to be n-dimensional. To facilitate the
calculation one can replace $\gamma_5$ in Eq. (\ref{eqn2.2})
by (see \cite{akde})
\begin{eqnarray}
\label{eqn2.22}
\gamma_5=\frac{i}{24}\,\epsilon_{\mu\nu\lambda\sigma}\,
\gamma^{\mu}\, \gamma^{\nu}\,\gamma^{\lambda}\,\gamma^{\sigma}\,,
\end{eqnarray}
which is equivalent to the HVBM scheme. Choosing the
${\overline{\rm MS}}$ subtraction scheme the renormalization constant
corresponding to the operator $O$ becomes \cite{klzu}
\begin{eqnarray}
\label{eqn2.23}
Z_O&=&1+a_s(\mu_r^2)\,S_{\varepsilon}\,\frac{2}{\varepsilon}\,\beta_0+
a_s^2(\mu_r^2)\,S_{\varepsilon}^2\,\Bigg [\frac{4}{\varepsilon^2}\,\beta_0^2
+\frac{2}{\varepsilon}\,\beta_1\Bigg ]+\cdots
\end{eqnarray}
where $S_{\varepsilon}$ denotes the spherical factor characteristic
of n-dimensional regularization. It is defined by
\begin{eqnarray}
\label{eqn2.24}
S_{\varepsilon}=\exp\left \{\frac{\varepsilon}{2}\Big [\gamma_E-\ln 4\pi\Big ]
\right \}\,.
\end{eqnarray}
The lowest order coefficients $\beta_0$ and $\beta_1$ originate from the
beta-function given by
\begin{eqnarray}
\label{eqn2.25}
&&\beta(\alpha_s)=-a_s^2(\mu_r^2)\,\beta_0-a_s^3(\mu_r^2)\,\beta_1+\cdots\,,
\nonumber\\[2ex]
&&\beta_0=\frac{11}{3}\,C_A-\frac{4}{3}\,n_f\,T_f \,, \quad
\beta_1=\frac{34}{3}\,C_A^2- 4\,n_f\,T_f\,C_F-\frac{20}{3}\,n_f\,T_f\,C_A\,.
\nonumber\\[2ex]
\end{eqnarray}
The operator renormalization constants corresponding to $O_1$ and $O_2$
are computed in \cite{larin} and they read
\begin{eqnarray}
\label{eqn2.26}
Z_{11}=Z_{\alpha_s}=1+a_s(\mu_r^2)\,S_{\varepsilon}\,\frac{2}{\varepsilon}\,
\beta_0+ a_s^2(\mu_r^2)\,S_{\varepsilon}^2\,\Bigg [\frac{4}{\varepsilon^2}\,
\beta_0^2
+\frac{1}{\varepsilon}\,\beta_1\Bigg ]+\cdots
\end{eqnarray}
where $Z_{\alpha_s}$ denotes the coupling constant renormalization factor
defined by
\begin{eqnarray}
\label{eqn2.27}
\hat a_s =Z_{\alpha_s}\,a_s(\mu_r^2)\,.
\end{eqnarray}
The remaining constants are
\begin{eqnarray}
\label{eqn2.28}
Z_{21}&=&0\,,
\\[2ex]
\label{eqn2.29}
Z_{12}&=&a_s(\mu_r^2)\,S_{\varepsilon}\,\frac{1}{\varepsilon}\,
\Big [-6\,C_F\Big ]\,,
\\[2ex]
\label{eqn2.30}
Z_{22}&=&Z^s_{\rm MS}\,Z_5^s\,,
\end{eqnarray}
where $Z^s_{\rm MS}$ and $Z_5^s$ are the constants characteristic of the
HVBM scheme. They are given by
\begin{eqnarray}
\label{eqn2.31}
Z^s_{\rm MS}&=&1+a_s^2(\mu_r^2)\,S_{\varepsilon}\,\frac{1}{\varepsilon}\,
\left [-\frac{44}{3}\,C_A\,C_F-n_f\,T_f\,C_F\,\frac{20}{3}\right ]\,,
\nonumber\\[2ex]
Z_5^s&=&1+a_s(\mu_r^2)\,\Big [-4\,C_F\Big ]+a_s^2(\mu_r^2)\,\left [22\,C_F^2
-\frac{107}{9}\,C_A\,C_F+\frac{31}{9}\,n_f\,T_f\,C_F\right ]\,.
\nonumber\\[2ex]
\end{eqnarray}
The latter renormalization constant is determined in such a way that
the Adler-Bell-Jackiw anomaly \cite{abj}
\begin{eqnarray}
\label{eqn2.32}
O_2(y)=4\,a_s(\mu_r^2)\,n_f\,T_f\,O_1(y)\,,
\end{eqnarray}
is preserved in all orders in perturbation theory according to the
Adler-Bardeen theorem \cite{adba}.

\mysection[3]{Two-loop vertex correction to the process $g + g \rightarrow H,A$.}
In this section we present the calculation of the (pseudo-) scalar Higgs
vertex corrected up to two loops. The scalar Higgs vertex was presented in 
\cite{harland} but no separation in colour factors was made. An explicit
formula for the pseudo-scalar vertex was not shown but it indirectly
appears in the coefficient functions of Higgs boson production
in \cite{haki3} and \cite{anme2}.

The one loop vertex correction is very easy. In the case of Higgs production
the characteristic graphs are shown in Fig. \ref{fig1}. For the operators
$O(y)$ and $O_1(y)$ all the solid lines represent gluons and there are only
two types of graphs.
The two-loop vertex correction, which is less trivial, can be calculated in
various ways. The characteristic graphs are shown in Fig. \ref{fig2}.
The solid lines represent gluons, quarks as well as ghosts (Feynman gauge).
Using the program FORM \cite{form} each graph can be reduced
into scalar integrals of the type
\begin{eqnarray}
\label{eqn3.1}
&&V^{ij}_{j_1j_2 \cdots j_n}=
\int\frac{d^nk_1}{(2\pi)^n}\,\int\frac{d^nk_2}{(2\pi)^n}
\frac{(k_l\cdot p_1)^i\,(k_m\cdot p_2)^j}{D_{j_1}\,D_{j_2}\cdots D_{j_n}}
\qquad l,m=1,2
\nonumber\\[2ex]
&&D_1=(k_1+k_2-p_1)^2\,,\quad D_2=(k_1+k_2-p_2)^2\,,\quad D_3=k_1^2\,,
\nonumber\\[2ex]
&&  D_4=(k_2-p_1)^2\,,\quad D_5=(k_2-p_2)^2\,,\quad D_6=k_2^2 \,,
\nonumber\\[2ex]
&& \quad D_7=(k_1-p_1)^2\,,\quad D_8=(k_1-p_2)^2\,.
\end{eqnarray}
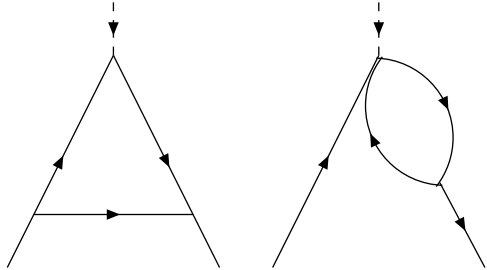
\begin{figure}[t]
\begin{center}
\begin{picture}(200,110)(0,0)
\ArrowLine(10,10)(50,90)
\ArrowLine(50,90)(90,10)
\ArrowLine(20,30)(80,30)
\DashArrowLine(50,110)(50,90){3}

\ArrowLine(110,10)(150,90)
\ArrowLine(173,42)(190,10)
\DashArrowLine(150,110)(150,90){3}

\ArrowArcn(148,59)(30,88,322)
\ArrowArcn(175,71)(30,268,142)

\end{picture}
\caption[]{General structure of the one-loop vertex correction g-g-B
($B=H,A$).}
\label{fig1}
\end{center}
\end{figure}
\begin{figure}
\begin{center}
\begin{picture}(360,120)(0,0)

\ArrowLine(10,10)(20,30)
\ArrowLine(100,30)(110,10)
\ArrowLine(20,30)(40,70)
\ArrowLine(80,70)(100,30)
\ArrowLine(40,70)(60,110)
\ArrowLine(60,110)(80,70)

\ArrowLine(20,30)(100,30)
\ArrowLine(40,70)(80,70)

\DashArrowLine(60,130)(60,110){3}

\ArrowLine(130,10)(140,30)
\ArrowLine(220,30)(230,10)

\ArrowLine(140,30)(160,70)
\ArrowLine(200,70)(220,30)

\ArrowLine(160,70)(180,110)
\ArrowLine(180,110)(200,70)
\ArrowLine(140,30)(200,70)
\ArrowLine(160,70)(220,30)

\DashArrowLine(180,130)(180,110){3}

\ArrowLine(250,10)(265,40)
\ArrowLine(265,40)(300,110)
\ArrowLine(300,110)(326,60)
\ArrowLine(326,60)(340,30)
\ArrowLine(340,30)(350,10)

\ArrowLine(265,40)(300,40)
\ArrowLine(300,40)(326,60)
\ArrowLine(340,30)(300,40)

\DashArrowLine(300,130)(300,110){3}

\end{picture}
\begin{picture}(360,120)(0,0)

\ArrowLine(10,10)(20,30)
\ArrowLine(100,30)(110,10)
\ArrowLine(20,30)(60,110)
\ArrowLine(60,110)(72,88)
\ArrowLine(90,50)(100,30)

\ArrowLine(20,30)(100,30)

\ArrowArcn(72,64)(25,88,322)
\ArrowArcn(90,73)(25,268,142)

\DashArrowLine(60,130)(60,110){3}

\ArrowLine(130,10)(145,40)
\ArrowLine(215,40)(230,10)

\ArrowLine(145,40)(180,110)
\ArrowLine(180,110)(215,40)

\ArrowLine(145,40)(155,40)
\ArrowLine(205,40)(215,40)

\ArrowArcn(180,30)(25,160,20)
\ArrowArcn(180,50)(25,340,200)

\DashArrowLine(180,130)(180,110){3}

\ArrowLine(250,10)(265,40)
\ArrowLine(265,40)(300,110)
\ArrowLine(300,110)(326,60)
\ArrowLine(326,60)(340,30)
\ArrowLine(340,30)(350,10)

\ArrowLine(265,40)(326,60)
\ArrowLine(340,30)(265,40)

\DashArrowLine(300,130)(300,110){3}

\end{picture}
\begin{picture}(360,120)(0,0)

\ArrowLine(10,10)(20,30)
\ArrowLine(100,30)(110,10)
\ArrowLine(20,30)(60,110)
\ArrowLine(81,70)(100,30)

\ArrowLine(20,30)(100,30)

\ArrowArcn(61,85)(25,88,322)
\ArrowArcn(81,94)(25,268,142)

\DashArrowLine(60,130)(60,110){3}

\ArrowLine(130,10)(140,30)
\ArrowLine(220,30)(230,10)
\ArrowLine(140,30)(180,110)
\ArrowLine(180,110)(200,70)

\ArrowLine(140,30)(220,30)

\ArrowArcn(200,47)(25,88,322)
\ArrowArcn(220,56)(25,268,142)

\DashArrowLine(180,130)(180,110){3}

\ArrowLine(250,10)(265,40)
\ArrowLine(335,40)(350,10)

\ArrowLine(265,40)(300,110)
\ArrowLine(300,110)(335,40)

\ArrowLine(265,40)(288,40)

\ArrowArcn(313,30)(25,160,20)
\ArrowArcn(313,50)(25,340,200)

\DashArrowLine(300,130)(300,110){3}

\end{picture}

\begin{picture}(360,100)(0,0)

\ArrowLine(10,0)(60,90)
\ArrowLine(60,90)(110,0)

\ArrowArcn(60,-5)(48,135,45)
\ArrowArcn(60,65)(48,315,225)

\DashArrowLine(60,110)(60,90){3}

\ArrowLine(130,0)(180,90)
\ArrowLine(180,90)(230,0)

\ArrowLine(140,15)(220,15)
\ArrowLine(180,90)(180,15)

\DashArrowLine(180,110)(180,90){3}

\ArrowLine(250,0)(259,15)

\ArrowLine(259,15)(300,90)

\ArrowLine(259,15)(338,15)
\ArrowLine(338,15)(348,0)

\ArrowArcn(289,40)(55,78,332)
\ArrowArcn(351,69)(55,258,152)

\DashArrowLine(300,110)(300,90){3}

\end{picture}
\end{center}
\end{figure}
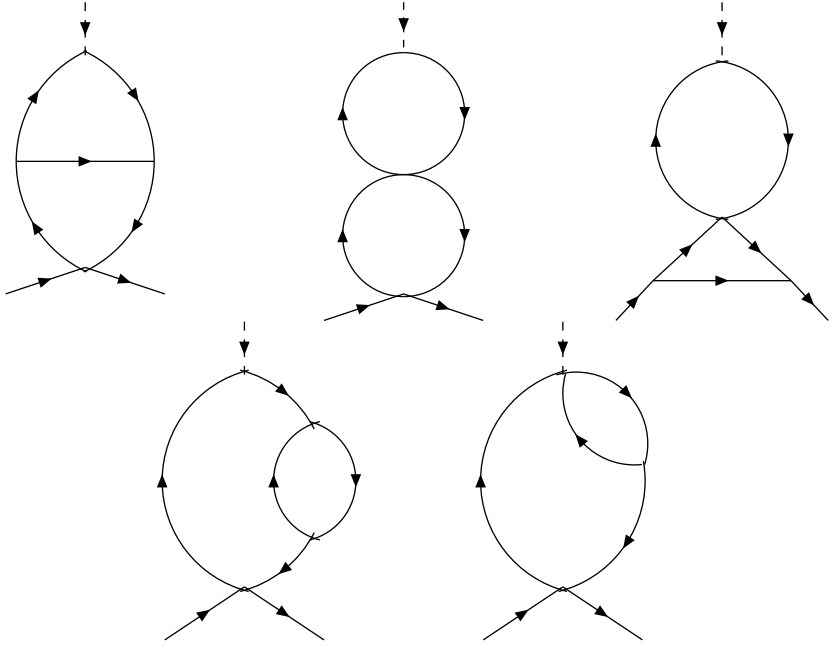
\begin{figure}[ht]
\begin{center}

\begin{picture}(360,120)(0,0)
\DashArrowLine(60,120)(60,100){3}
\ArrowArcn(40,60)(46,65,0)
\ArrowArcn(40,60)(46,0,295)
\ArrowArcn(80,60)(46,245,180)
\ArrowArcn(80,60)(46,180,115)

\ArrowLine(30,10)(60,20)
\ArrowLine(60,20)(90,10)

\ArrowLine(34,60)(86,60)

\DashArrowLine(180,120)(180,103){3}

\ArrowArcn(180,78)(23,90,270)
\ArrowArcn(180,78)(23,270,90)
\ArrowArcn(180,32)(23,90,270)
\ArrowArcn(180,32)(23,270,90)

\ArrowLine(150,0)(180,10)
\ArrowLine(180,10)(210,0)

\DashArrowLine(300,120)(300,100){3}

\ArrowArcn(295,68)(30,85,275)
\ArrowArcn(305,68)(30,265,95)

\ArrowLine(274,15)(300,39)
\ArrowLine(300,39)(326,15)

\ArrowLine(260,0)(274,15)
\ArrowLine(326,15)(340,0)

\ArrowLine(274,15)(326,15)

\end{picture}
\begin{picture}(360,120)(0,0)

\DashArrowLine(120,120)(120,100){3}
\ArrowArcn(108,60)(43,76,27)
\ArrowArcn(108,60)(43,333,284)
\ArrowArcn(132,60)(43,256,104)

\ArrowArcn(139,60)(23,76,284)
\ArrowArcn(154,60)(23,256,104)

\ArrowLine(90,0)(120,20)
\ArrowLine(120,20)(150,0)

\DashArrowLine(240,120)(240,100){3}
\ArrowArcn(228,60)(43,10,284)
\ArrowArcn(252,60)(43,256,104)

\ArrowArcn(245,74)(27,106,343)
\ArrowArcn(267,93)(27,276,164)

\ArrowLine(210,0)(240,20)
\ArrowLine(240,20)(270,0)

\end{picture}

\caption[]{General structure of the two-loop vertex correction g-g-B
($B=H,A$).}
\label{fig2}
\end{center}
\end{figure}
The scalar integrals in the case of the Higgs boson vertex are ordered
as follows. For six propagators we have $i+j=4$. For five propagators $i+j=3$.
For four propagators $i+j=2$. For three propagators $i+j=1$. The two and
one propagator integrals are all zero.
The calculation of these scalar integrals can be performed in various ways.
For the electromagnetic vertex correction in QCD it has been done in
\cite{gon}, \cite{krla}, \cite{mane}. In \cite{gon} one has chosen a
Feynman parameter set in a judicious way so that after each integration
one is still left with a linear integral over the next parameter.
In \cite{krla} one used integration by parts according to \cite{chtk}.
This trick eliminates one propagator from the integral in Eq. (\ref{eqn3.1}).
However it does not work for the second Feynman graph in Fig. \ref{fig2}
(the crossed box graph).
In \cite{mane} the scalar integrals were calculated using dispersion techniques
developed in \cite{ne}. In the latter reference one has used the Cutkosky
rules \cite{cut} to compute the scalar integrals. Here one cuts the graph
in all possible ways to obtain the imaginary part. The real part is
obtained via a dispersion relation. Fortunately for the electromagnetic
vertex correction $i+j\le 3$ in Eq. (\ref{eqn3.1}) but for the vertex 
correction
to $O(y)$ and $O_1(y)$ in Eq. (\ref{eqn2.2}) one also gets contributions for 
$i+j=4$. In \cite{harland} one employed an algorithm in \cite{basm} which 
relates
l-loop integrals with $n+1$ external legs to $l+1$-loop integrals with n
external legs. In the case under consideration massless two-loop vertex
functions are mapped onto massless three-loop two-point functions.
We however adopt the method in \cite{ne} and calculate the imaginary
part of all graphs using the Cutkosky rules. In the case that all particles
are massless, except for the external Higgs boson leg,
the dispersion integrals are trivial.
Since all graphs, except for the second graph in Fig. \ref{fig2},
are analytic we have made a computer program in FORM \cite{form} which
provides us with all scalar integrals of the type in (\ref{eqn3.1}). For more 
details about the real and imaginary parts of the scalar integrals we refer
to \cite{ne}. The result for the scalar Higgs vertex is obtained from 
Eq. (\ref{eqn2.19}) and it reads up to order $\hat a_s^2$
\begin{eqnarray}
\label{eqn3.2}
F_H(q^2)\,(2\pi)^4\,\delta(q+p_1-p_2)=Z_O\,\int d^4y\,e^{i\,q\cdot y}\,
\langle g(p_2) |\hat O(y)| g(p_1) \rangle \,,
\end{eqnarray}
where $F_H(q^2)$ for $q^2<0$ is equal to ($n=4+\varepsilon$)
\begin{eqnarray}
\label{eqn3.3}
Z_O^{-1}\,F_H(q^2)&=&1+\hat a_s\,S_\varepsilon\,\left (\frac{-q^2}{\mu^2} 
\right )
^{\varepsilon/2}\,
C_A\,\Bigg [-\frac{8}{\varepsilon^2}+\zeta(2)+\varepsilon\,\Big (1-\frac{7}{3}
\,\zeta(3)\Big )
\nonumber\\[2ex]
&&+\varepsilon^2\,\Big (-\frac{3}{2}+\frac{47}{80}\,\zeta^2(2)\Big )\Bigg ]
+\hat a_s^2\,S_\varepsilon^2\, \left (\frac{-q^2}{\mu^2} 
\right )^{\varepsilon}\,\Bigg [C_A^2\,\Bigg \{
\frac{32}{\varepsilon^4}+\frac{44}{3\,\varepsilon^3}
\nonumber\\[2ex]
&&-\Big (\frac{134}{9}
+4\,\zeta(2)\Big )\,\frac{1}{\varepsilon^2}
+\Big (-\frac{136}{27}-11\,\zeta(2)
+\frac{50}{3}\,\zeta(3)\Big )\,\frac{1}{\varepsilon}
\nonumber\\[2ex]
&&+\frac{5861}{162}
+\frac{67}{6}\,\zeta(2)+\frac{11}{9}\,\zeta(3)
-\frac{21}{5}\,\zeta^2(2)\Bigg \}
\nonumber\\[2ex]
&&+n_f\,T_f\,C_A\,\Bigg \{-\frac{16}{3\,\varepsilon^3}+
\frac{40}{9\,\varepsilon^2}
+\Big (\frac{104}{27}+4\,\zeta(2)\Big )\,\frac{1}{\varepsilon}-\frac{1616}{81}
\nonumber\\[2ex]
&&-\frac{10}{3}\,\zeta(2)-\frac{148}{9}\,\zeta(3)\Bigg \}
\nonumber\\[2ex]
&&+ n_f\,T_f\,C_F\,\Bigg \{\frac{4}{\varepsilon}-\frac{67}{3}+16\,\zeta(3)
\Bigg \} \Bigg ]\,,
\end{eqnarray}
where $\hat a_s$ (see Eq. (\ref{eqn2.4})) is the bare coupling.
The result above agrees with Eqs. (16) and (17) in \cite{harland} 
provided we put $C_A=3$ and $C_F=4/3$ (Notice
that in \cite{harland} $n=4-2\,\varepsilon$).
The pseudo-scalar Higgs vertex can be derived from Eq. (\ref{eqn2.20}).
\begin{eqnarray}
\label{eqn3.4}
F_A(q^2)\,(2\pi)^4\,\delta(q+p_1-p_2)&=&\int d^4y\,e^{i\,q\cdot y}\,
\Bigg \{Z_{11}\,\langle g(p_2) |\hat O_1(y)| g(p_1) \rangle
\nonumber\\[2ex]
&&+Z_{12}\,\langle g(p_2) |\hat O_2(y)| g(p_1) \rangle\Bigg \}\,,
\end{eqnarray}
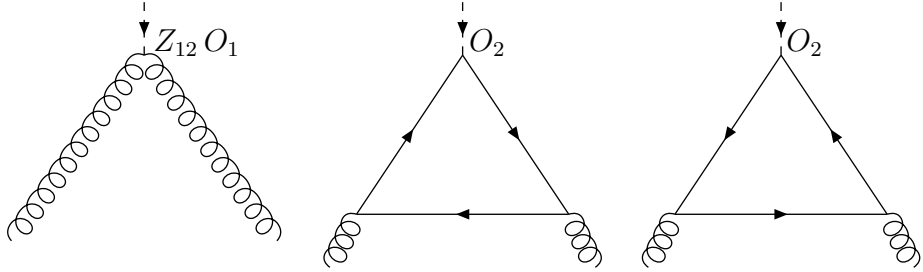
\begin{figure}
\begin{center}
\begin{picture}(360,100)(0,0)
\DashArrowLine(60,100)(60,80){3}
\Gluon(10,10)(60,80){4}{11}
\Gluon(60,80)(110,10){4}{11}

\DashArrowLine(180,100)(180,80){3}

\Gluon(130,0)(140,20){4}{3}
\Gluon(220,20)(230,0){4}{3}
\ArrowLine(140,20)(180,80)
\ArrowLine(180,80)(220,20)
\ArrowLine(220,20)(140,20)

\DashArrowLine(300,100)(300,80){3}

\Gluon(250,0)(260,20){4}{3}
\Gluon(340,20)(350,0){4}{3}
\ArrowLine(300,80)(260,20)
\ArrowLine(340,20)(300,80)
\ArrowLine(260,20)(340,20)

\Text(80,90)[t]{$Z_{12}\,O_1$}
\Text(190,90)[t]{$O_2$}
\Text(310,90)[t]{$O_2$}

\end{picture}
\caption[]{Interference term between diagrams with operators
$O_1$ and $O_2$.}
\label{fig3}
\end{center}
\end{figure}
The matrix element $\langle g(p_2) |\hat O_2(y)| g(p_1) \rangle=C\,
\hat \alpha_s$
but $Z_{12}\sim \hat \alpha_s\,\varepsilon^{-1}$ (see Eq. (\ref{eqn2.29})).
This contribution has to be added because we use the HVBM scheme for 
the $\gamma_5$-matrix. The extra term equals (see Fig. \ref{fig3})
\begin{eqnarray}
\label{eqn3.5}
\hat a_s^2\,S_\varepsilon^2\,C_F\,T_f\,\left (\frac{-q^2}{\mu^2} \right )
^{\varepsilon}\,\Bigg [\frac{24}{\varepsilon}-36\Bigg ]\,.
\end{eqnarray}
In the equation above we have multiplied by an extra factor $S_\varepsilon\,
(-q^2/\mu^2)^{\varepsilon/2}$ to restore the 
anti-commutativity of the $\gamma_5$-matrix. In a scheme where the
$\gamma_5$-matrix anti-commutes
with the other gamma matrices this extra factor would be automatically there.
This term has to be added to $F_A(q^2)$ and the result up to
order $\hat a_s^2$ becomes ($q^2<0$)
\begin{eqnarray}
\label{eqn3.6}
Z_{11}^{-1}\,F_A(q^2)&=&1+\hat a_s\,S_\varepsilon\,
\left (\frac{-q^2}{\mu^2} \right )
^{\varepsilon/2}\,
C_A\,\Bigg [-\frac{8}{\varepsilon^2}+4+\zeta(2)+\varepsilon\,\Big (-6
-\frac{7}{3}\,\zeta(3)\Big )
\nonumber\\[2ex]
&&+\varepsilon^2\,\Big (7-\frac{1}{2}\,\zeta(2)+\frac{47}{80}\,\zeta^2(2)
\Big )\Bigg ]+\hat a_s^2\,S_\varepsilon^2\,\left (\frac{-q^2}{\mu^2}
\right )^{\varepsilon}\,C_A^2\,\Bigg [
\frac{32}{\varepsilon^4}
\nonumber\\[2ex]
&&+\frac{44}{3\,\varepsilon^3}-\Big (\frac{422}{9} 
+4\,\zeta(2)\Big )\,\frac{1}{\varepsilon^2}
+\Big (\frac{890}{27}-11\,\zeta(2)
+\frac{50}{3}\,\zeta(3)\Big )\,\frac{1}{\varepsilon}
\nonumber\\[2ex]
&&+\frac{3835}{81}+\frac{115}{6}\,\zeta(2)+\frac{11}{9}\,\zeta(3)
-\frac{21}{5}\,\zeta^2(2)\Bigg ]
\nonumber\\[2ex]
&&+n_f\,T_f\,C_A\,\Bigg \{-\frac{16}{3\,\varepsilon^3}
+\frac{40}{9\,\varepsilon^2}
+\Big (\frac{212}{27}+4\,\zeta(2)\Big )\,\frac{1}{\varepsilon}
-\frac{3182}{81}
\nonumber\\[2ex]
&&-\frac{10}{3}\,\zeta(2)-\frac{148}{9}\,\zeta(3)\Bigg \}
\nonumber\\[2ex]
&&+ n_f\,T_f\,C_F\,\Bigg \{-\frac{142}{3}+16\,\zeta(3) \Bigg \} \Bigg ]\,.
\end{eqnarray}
This result is new. It only indirectly appears in the $\delta(1-z)$ part
of the total cross section for pseudo-scalar Higgs boson production
in \cite{haki3} and \cite{anme2}. We can now compute the part due
to virtual and soft gluons which is proportional to $\delta(1-z)$
for general colour factors as in \cite{rasm2}. Continuing the above 
form factors to $q^2>0$ and adding the soft gluon cross section we 
can write the soft plus virtual gluon contribution to the coefficient 
function in Eq. (\ref{eqn2.11}) as
\begin{eqnarray}
\label{eqn3.7}
\Delta_{gg,{\rm B}}^{\rm S+V}=\delta(1-z)+a_s\,
\Delta_{gg,{\rm B}}^{(1),{\rm S+V}}+a_s^2\,\Delta_{gg,{\rm B}}^{(2),{\rm S+V}}
\,.
\end{eqnarray}
Starting with scalar Higgs boson production we omit the term which
is proportional to $C_A^2$ as it is available in \cite{rasm2}. 
The remaining order $a_s^2$ part of Eq. (\ref{eqn3.7}) equals  
\begin{eqnarray}
\label{eqn3.8}
\Delta_{gg,{\rm H}}^{(2),{\rm S+V}}&=&
n_f\,T_f\,C_A\,\Bigg [\Bigg \{\frac{16}{3}\,{\cal D}_0(z)\Bigg \}
\ln^2 \left(\frac{m^2}{\mu^2}\right)+\Bigg \{\frac{64}{3}\,{\cal D}_1(z)
- \frac{160}{9}\,{\cal D}_0(z)
\nonumber\\[2ex]
&& +\delta(1-z)\Big (16+\frac{32}{3}\,\zeta(2)\Big )\Bigg \}
\,\ln \left(\frac{m^2}{\mu^2}\right)+\frac{64}{3}\,{\cal D}_2(z)
-\frac{320}{9}\,{\cal D}_1(z)
\nonumber\\[2ex]
&&+\Big (\frac{448}{27}
-\frac{64}{3}\,\zeta(2)\Big )\,{\cal D}_0(z)+\delta(1-z)\,\Big (
- \frac{160}{3}-\frac{160}{9}\,\zeta(2)
\nonumber\\[2ex]
&&-\frac{16}{3}\, \zeta(3)\Big )\Bigg ]
\nonumber\\[2ex]
&& + n_f\,T_f\,C_F\,\delta(1-z)\,\Bigg [8\,\ln \left(\frac{m^2}{\mu^2}
\right) -\frac{134}{3}+32\,\zeta(3)\Bigg ]\,.
\end{eqnarray}
The result above is in agreement with \cite{haki3} and \cite{anme1}
for $C_A=3$ and $C_F=4/3$.
For the pseudo scalar case we take the difference with the scalar Higgs
boson coefficient function. Moreover we add the last term in 
Eq. (\ref{eqn2.20}) which equals
\begin{eqnarray}
\label{eqn3.9}
 \int d^4y\,e^{i\,q\cdot y}\,G_{\rm A}\,\tilde G_{\rm A}\,
\langle g,g|\hat O_1(y)\,\hat O_2(0) +\hat O_2(y)\,\hat O_1(0)|g,g\rangle
\,.
\end{eqnarray} 
This term is proportional to $O_{12}$ in the formula below and it originates
from Eq. (\ref{eqn2.3}). Ignoring the term proportional to $C_A^2$ which is 
given in \cite{rasm2} the remaining order $a_s^2$ term equals
\begin{eqnarray}
\label{eqn3.10}
\Delta_{gg,{\rm A-H}}^{(2),{\rm S+V}}&=&
C_A\,T_f\,n_f\,\delta(1-z)\,\Bigg [-\frac{8}{3}\,
\ln \left(\frac{m^2}{\mu^2}\right) -\frac{4}{3}\Bigg ]
\nonumber\\[2ex]
&&+C_F\,T_f\,n_f\,\delta(1-z)\,\Bigg [-8\,\ln \left(\frac{m^2}{\mu^2}
\right)-50
\nonumber\\[2ex]
&&+O_{12}\,\left (24\,\ln \left(\frac{\mu_r^2}{m_t^2}
\right)-12\right ) \Bigg ]\,,
\end{eqnarray}
which agrees with \cite{haki3} and \cite{anme2} for $C_A=3$ and $C_F=4/3$.

\mysection{General structure of the vertex correction}
In this section we will study the general structure of the vertex correction
as predicted in \cite{sen} and \cite{col}. In these two papers the behaviour
of the vertex corrections w.r.t. $\ln (-q^2/\mu^2)$ was derived. Following
Eq. (6.21) in \cite{col} the behaviour is as follows
\begin{eqnarray}
\label{eqn4.1}
\frac{d\,\ln F_i(q^2)}{d\,\ln (-q^2/\mu^2)}&=&-\frac{1}{4}\,\int_{\mu^2}^{-q^2}
\frac{d\,\mu_1^2}{\mu_1^2}\,\gamma_{K,ii}\Big (a_s(\mu_1^2)\Big )
+\frac{1}{2}\,G_{ii}\Big (a_s(q^2)\Big )
\nonumber\\[2ex]
&&+K_{ii}\left (\frac{1}{\varepsilon},a_s(\mu^2)\right )\,,
\nonumber\\
\end{eqnarray}
where 
\begin{eqnarray}
\label{eqn4.2}
\gamma_{K,ii}=a_s(\mu^2)\,\gamma_{K,ii}^{(0)}+a_s^2(\mu^2)\,
\gamma_{K,ii}^{(1)}+a_s^3(\mu^2)\,
\gamma_{K,ii}^{(2)}+ \cdots
\end{eqnarray}
and
\begin{eqnarray}
\label{eqn4.3}
G_{ii}&=&a_s(q^2)\,\Big [\gamma_{ii}^{(0)}-z_1\Big ]
+a_s^2(q^2)\,\Big [\gamma_{ii}^{(1)}-z_2\Big ]
\nonumber\\[2ex]
&&+a_s^3(q^2)\,\Big [
\gamma_{ii}^{(2)}-z_3\Big ]+ \cdots
\end{eqnarray}
The quantities $\gamma_{K,ii}$ and $G_{ii}$
($i=q,g$) occur in the diagonal kernel
\begin{eqnarray}
\label{eqn4.4}
\Gamma_{ii}&=&1+\Bigg [\gamma_{K,ii}\,\left (\frac{1}{1-x}\right )_+
+G_{ii}\,\delta(1-x)\Bigg ]\,,
\end{eqnarray}
and they represent those parts of the splitting functions which are
proportional to the distributions $(1/(1-x))_+$ and 
$\delta(1-x)$ respectively provided they are expressed in the
${\overline {\rm MS}}$-scheme. They have been calculated up to three loop
order in \cite{move}. The constants $z_i$ occur when the operator
is not conserved which happens in the case of Higgs boson vertex corrections.
The constants show up in the operator renormalization constants $Z_O$ 
in Eq. (\ref{eqn2.23}) and $Z_{11}$ in Eq. (\ref{eqn2.26}).  
The expression for these operator renormalization constants is ($Z=Z_O$
or $Z=Z_{11}$)
\begin{eqnarray}
\label{eqn4.5}
Z=1+a_s\,S_\varepsilon\,\Bigg [\frac{z_1}{\varepsilon}\Bigg ]+
a_s^2\,S_\varepsilon^2\,\Bigg [\frac{1}{\varepsilon^2}\,\Big (\frac{1}{2}\,
z_1^2+\beta_0\,z_1\Big )+ \frac{z_2}{2\,\varepsilon}\Bigg ]\,.
\end{eqnarray}
Here $z_1=2\,\beta_0$ for both operators. Further we have $z_2=4\,\beta_1$
for $O(y)$ in Eq. (\ref{eqn2.23}) and
$z_2=2\,\beta_1$ for $O_1(y)$ in Eq. (\ref{eqn2.26}).
If we express this in the bare coupling constant it becomes
\begin{eqnarray}
\label{eqn4.6}
Z=1+\hat a_s\,S_\varepsilon\,\Bigg [\frac{z_1}{\varepsilon}\Bigg ]+
\hat a_s^2\,S_\varepsilon^2\,\Bigg [\frac{1}{\varepsilon^2}\,\Big (\frac{1}{2}
\,z_1^2 -\beta_0\,z_1\Big )+\frac{z_2}{2\,\varepsilon}\Bigg ]\,.
\end{eqnarray}
In the case of the electromagnetic vertex the current is conserved and we
find $z_i=0$.
Finally $K_{ii}$ in Eq. (\ref{eqn4.1}) collects all collinear and 
infrared terms.
Because of the one to one correspondence between the $\ln (-q^2/\mu^2)$
terms and the pole terms $1/\varepsilon$ it reads up to order $a_s^2$
\begin{eqnarray}
\label{eqn4.7}
K_{ii}&=&a_s(\mu^2)\,S_\varepsilon\,\Bigg [-\frac{1}{2\,\varepsilon}\,
\gamma_{K,ii}^{(0)}\Bigg ]+a_s^2(\mu^2)\,S_\varepsilon^2\,\Bigg [
-\frac{1}{2\,\varepsilon^2}\,\beta_0\,\gamma_{K,ii}^{(0)}-
\frac{1}{4\,\varepsilon}\,\gamma_{K,ii}^{(1)}+f_{i,2}^{(1)}
\nonumber\\[2ex]
&&-\beta_0\,f_{i,1}^{(0)}
\Bigg ]+\cdots
\end{eqnarray}
where $f_1^{(0)}$ is a vertex dependent constant. However $f_{i,2}^{(1)}$
is vertex independent as we shall see later on.
From Eq. (\ref{eqn4.1}) one can derive the formula for the logarithm
of the vertex correction. First we replace the coupling constant
$a_s(\mu_1^2)$ by $a_s(\mu^2)$
\begin{eqnarray}
\label{eqn4.8}
a_s(\mu_1^2)=a_s(\mu^2)\,\Bigg (1-a_s(\mu^2)\,\beta_0\,
\ln \frac{\mu_1^2} {\mu^2} \Bigg )\,,
\end{eqnarray}
then we integrate over $\mu_1^2$. Subsequently we integrate over 
$\ln (-q^2/\mu^2)$ and finally we replace the renormalized by the bare coupling
constant
\begin{eqnarray}
\label{eqn4.9}
a_s(\mu^2)=\hat a_s\,\Bigg (1-\hat a_s\,\frac{2}{\varepsilon}\,S_\varepsilon\,
\beta_0\Bigg )\,.
\end{eqnarray}
We can now integrate Eq. (\ref{eqn4.1}) over $\ln (-q^2/\mu^2)$. 
However in the case of non 
conserved operators one has to be careful in the determination of
the pole part of the vertex function. First we separate the 
$\ln (-q^2/\mu^2)$ terms which are due to
the non conserved operators from the rest. Next we observe a one
to one correspondence between the pole part and the remaining 
$\ln (-q^2/\mu^2)$ piece. Hence the general expression for the unrenormalized 
vertex correction becomes
\begin{eqnarray}
\label{eqn4.10}
\ln F_i(q^2)&=&\hat a_s\,S_\varepsilon\,\left (\frac{-q^2}{\mu^2} \right )
^{\varepsilon/2}\,\Bigg [
-\frac{\gamma_{K,ii}^{(0)}}{\varepsilon^2}+\frac{\gamma_{ii}^{(0)}}
{\varepsilon} +f_{i,1}^{(0)}\Bigg ]
\nonumber\\[2ex]
&&-\hat a_s\,S_\varepsilon\,\Bigg [\frac{1}{2}\,z_1\,\ln \left (\frac{-q^2}{\mu^2}
\right ) +\frac{1}{8}\,\varepsilon\,z_1\,\ln^2\left ( \frac{-q^2}{\mu^2}\right )\Bigg ]
\nonumber\\[2ex]
&&+\hat a_s^2\,S_\varepsilon^2\,\left (\frac{-q^2}{\mu^2} \right )
^{\varepsilon}\,\Bigg [
\frac{1}{2}\,\beta_0\,\gamma_{K,ii}^{(0)}\,\frac{1}{\varepsilon^3}
-\Big (\frac{1}{4}\,\gamma_{K,ii}^{(1)}+\beta_0\,\gamma_{ii}^{(0)}\Big )\,
\frac{1}{\varepsilon^2}
\nonumber\\[2ex]
&&+\Big (\frac{1}{2}\,\gamma_{ii}^{(1)}+f_{i,2}^{(1)}
-2\,\beta_0\,f_{i,1}^{(0)}\Big ) \,\frac{1}{\varepsilon}
+f_{i,2}^{(0)}\Bigg ]
\nonumber\\[2ex]
&&-\hat a_s^2\,S_\varepsilon^2\,\Bigg [
-\frac{\beta_0\,z_1}{\varepsilon}\,\ln \left (\frac{-q^2}{\mu^2}\right )
-\frac{1}{2}\,\beta_0\,z_1\,\ln^2 \left (\frac{-q^2}{\mu^2}\right )
\nonumber\\[2ex]
&& +\frac{1}{2}\,z_2\,\ln \left (\frac{-q^2}{\mu^2}\right )\Bigg ]\,,
\qquad i=H,A,\gamma^*\,.
\end{eqnarray}
Note that $z_k$ are implicit in $\gamma_{ii}^{(k-1)}$ and they do not
contribute to the $\ln (-q^2/\mu^2)$ terms. They contribute to the pole terms
$(1/\varepsilon)^k$ only. Another remark concerns the term 
$-2\,\beta_0\,f_{i,1}^{(0)}$ in the penultimate line of the above equation.
It vanishes in the single pole term after coupling constant renormalization.
The anomalous dimensions $\gamma_{K,ii}^{(k)}$ are up to a colour factor $C_i$
vertex independent. Up to
two-loops they are given by
\begin{eqnarray}
\label{eqn4.11}
\gamma_{K,ii}^{(0)}&=&8\,C_i\,, \quad \gamma_{K,ii}^{(1)}=16\,C_i\,K\,,
\quad K=-\frac{10}{9}\,n_f\,T_f+C_A\,\Big (\frac{67}{18}-\zeta(2)\Big )\,,
\nonumber\\[2ex]
i&=&q,g \,,\qquad C_q=C_F \,,\qquad C_g=C_A \,.
\end{eqnarray}
However, and this is very important, $f_{i,2}^{(1)}$ is also vertex independent
up to a colour factor $C_i$ and it is given by
\begin{eqnarray}
\label{eqn4.12}
f_{i,2}^{(1)}&=&C_i\,\Bigg [-\beta_0\,\zeta(2)
+C_A\,\Big (\frac{404}{27}-14\,\zeta(3)\Big )
+n_f\,T_f\,\Big (-\frac{112}{27}\Big )\Bigg ]\,.
\end{eqnarray}
The whole vertex dependence can be attributed to the the anomalous
dimensions $\gamma_{ii}^{(k)}$ and the non-pole terms $f_{i,k}^{(0)}$.
The anomalous dimensions up to two loops are given by
\begin{eqnarray}
\label{eqn4.13}
\gamma_{qq}^{(0)}=6\,C_F\,, \qquad \gamma_{gg}^{(0)}=2\,\beta_0\,,
\end{eqnarray}
\begin{eqnarray}
\label{eqn4.14}
\gamma_{qq}^{(1)}&=&C_F^2\,\Big (3-24\,\zeta(2)+48\,\zeta(3)\Big )+C_A\,C_F\,
\Big (\frac{17}{3}+\frac{88}{3}\,\zeta(2)-24\,\zeta(3)\Big )
\nonumber\\[2ex]
&&+n_f\,T_f\,C_F\,\Big (-\frac{4}{3}-\frac{32}{3}\,\zeta(2) \Big )\,,
\nonumber\\[2ex]
\gamma_{gg}^{(1)}&=&C_A^2\,\Big (\frac{64}{3}+24\,\zeta(3)\Big )-\frac{32}{3}
\,n_f\,T_f\,C_A-8\,n_f\,T_f\,C_F\,.
\end{eqnarray}
The non-pole terms $f_{i,k}^{(0)}$ ($i=q,g$) can be extracted from the actual
calculations. For that purpose we also include the 
quark vertex correction in the case of electromagnetism \cite{gon}, \cite{krla},
\cite{mane}. Starting with the latter, this expression equals
\begin{eqnarray}
\label{eqn4.15}
\ln F_{\gamma^*}(q^2)&=&\hat a_s\,S_\varepsilon\,\left (\frac{-q^2}{\mu^2}
\right ) ^{\varepsilon/2}\,C_F
\,\Bigg [-\frac{8}{\varepsilon^2}+\frac{6}{\varepsilon}-8+\zeta(2)\Bigg ]
\nonumber\\[2ex]
&&+\hat a_s^2\,S_\varepsilon^2\,\left (\frac{-q^2}{\mu^2} \right )
^{\varepsilon} \, \Bigg [C_F^2\,
\Bigg \{\Big ( \frac{3}{2}-12\,\zeta(2)+24\,\zeta(3)\Big )\,
\frac{1}{\varepsilon}-\frac{1}{8}+29\,\zeta(2)
\nonumber\\[2ex]
&&-30\,\zeta(3)-\frac{44}{5}\,
\zeta^2(2)\Bigg \}+C_A\,C_F\,\Bigg \{\frac{44}{3\,\varepsilon^3}+\Big (
-\frac{332}{9}+4\,\zeta(2)\Big )\,\frac{1}{\varepsilon^2}
\nonumber\\[2ex]
&&+\Big (\frac{4129}{54}+\frac{11}{3}\,\zeta(2)-26\,\zeta(3)\Big )
\,\frac{1}{\varepsilon}-\frac{89173}{648}-\frac{119}{9}\,\zeta(2)
\nonumber\\[2ex]
&&+\frac{467}{9}\,\zeta(3)+\frac{44}{5}\,\zeta^2(2)
\Bigg \}+n_f\,T_f\,C_F\,\Bigg \{-\frac{16}{3\,\varepsilon^3}
+\frac{112}{9}\,\frac{1}{\varepsilon^2}
\nonumber\\[2ex]
&&+\Big (-\frac{706}{27}-\frac{4}{3}\,
\zeta(2)\Big )\,\frac{1}{\varepsilon}+\frac{7541}{162}
+\frac{28}{9}\,\zeta(2)
\nonumber\\[2ex]
&&-\frac{52}{9}\,\zeta(3) \Bigg \}\Bigg ]\,.
\end{eqnarray}
The non-pole terms $f_{q,k}^{(0)}$ are
\begin{eqnarray}
\label{eqn4.16}
f_{q,1}^{(0)}&=&C_F\,(\zeta(2)-8)\,,
\nonumber\\[2ex]
f_{q,2}^{(0)}&=&C_F^2\,\Big (-\frac{1}{8}+29\,\zeta(2)-\frac{44}{5}\,\zeta^2(2)
-30\,\zeta(3)\Big )
+C_A\,C_F\,\Big (-\frac{89173}{648}
\nonumber\\[2ex]
&&-\frac{119}{9}\,\zeta(2)+\frac{44}{5}\,
\zeta^2(2)+\frac{467}{9}\,\zeta(3)\Big )
+n_f\,T_f\,C_F\,\Big (\frac{7541}{162}+\frac{28}{9}\,\zeta(2)
\nonumber\\[2ex]
&&-\frac{52}{9}\,\zeta(3)\Big )\,.
\end{eqnarray}
Next we turn our attention to Higgs boson production.
Taking the logarithm of the expression for the scalar Higgs boson Eqs. (\ref{eqn3.3})
we obtain 
\begin{eqnarray}
\label{eqn4.17}
\ln F_H(q^2)&=&\ln Z_O+\hat a_s\,S_\varepsilon\,\left (\frac{-q^2}{\mu^2} 
\right )^{\varepsilon/2}\,C_A
\,\Bigg [-\frac{8}{\varepsilon^2}+\zeta(2)\Bigg ]
\nonumber\\[2ex]
&&+\hat a_s^2\,S_\varepsilon^2\,\left (\frac{-q^2}{\mu^2} \right )^{\varepsilon}
\, \Bigg [C_A^2\,
\Bigg \{\frac{44}{3\,\varepsilon^3}+\Big (-\frac{134}{9}+4\,\zeta(2)\Big )
\,\frac{1}{\varepsilon^2}+\Big ( \frac{80}{27}
\nonumber\\[2ex]
&&-11\,\zeta(2)-2\,\zeta(3)\Big )\,
\frac{1}{\varepsilon}+\frac{3917}{162}+\frac{67}{6}\,\zeta(2)
+\frac{11}{9}\,\zeta(3)\Bigg \}
\nonumber\\[2ex]
&&+n_f\,T_f\,C_A\,\Bigg \{-\frac{16}{3\,\varepsilon^3}
+\frac{40}{9\,\varepsilon^2}
+\Big (\frac{104}{27}+4\,\zeta(2)\Big )\,\frac{1}{\varepsilon}
-\frac{1616}{81} 
\nonumber\\[2ex]
&&-\frac{10}{3}\,\zeta(2)-\frac{148}{9}\,\zeta(3)\Bigg \}
+n_f\,T_f\,C_F\,\Bigg \{\frac{4}{\varepsilon}
-\frac{67}{3}+16\,\zeta(3) \Bigg \}\Bigg ]\,,
\nonumber\\
\end{eqnarray}
with the non-pole constants $f_{g,k}^{(0)}$
\begin{eqnarray}
\label{eqn4.18}
f_{g,1}^{(0)}&=&C_A\,\zeta(2)\,,
\nonumber\\[2ex]
f_{g,2}^{(0)}&=&n_f\,T_f\,C_A\,\Big (-\frac{1616}{81}-\frac{10}{3}\,\zeta(2)
-\frac{148}{9}\,\zeta(3)\Big )
+n_f\,T_f\,C_F\,\Big (-\frac{67}{3}
\nonumber\\[2ex]
&&+16\,\zeta(3)\Big ) +C_A^2\,\Big
(\frac{3917}{162}+\frac{67}{6}\,\zeta(2)+\frac{11}{9}\,\zeta(3)\Big )\,.
\end{eqnarray}
Doing the same for the pseudo-scalar Higgs boson vertex (see Eq. (\ref{eqn3.6})) we get
\begin{eqnarray}
\label{eqn4.19}
\ln F_A(q^2)&=&\ln Z_{11}+\hat a_s\,S_\varepsilon\,\left (\frac{-q^2}{\mu^2} 
\right )^{\varepsilon/2}\,C_A
\,\Bigg [-\frac{8}{\varepsilon^2}+4+\zeta(2)\Bigg ]
\nonumber\\[2ex]
&&+\hat a_s^2\,S_\varepsilon^2\,\left (\frac{-q^2}{\mu^2} \right )^{\varepsilon}
\, \Bigg [C_A^2\,
\Bigg \{\frac{44}{3\,\varepsilon^3}+\Big (-\frac{134}{9}+4\,\zeta(2)\Big )
\,\frac{1}{\varepsilon^2}+\Big ( -\frac{406}{27}
\nonumber\\[2ex]
&&-11\,\zeta(2)-2\,\zeta(3)\Big )
\,\frac{1}{\varepsilon}+\frac{7723}{81}+\frac{67}{6}\,\zeta(2)
+\frac{11}{9}\,\zeta(3)\Bigg \}
\nonumber\\[2ex]
&&+n_f\,T_f\,C_A\,\Bigg \{-\frac{16}{3\,\varepsilon^3}
+\frac{40}{9\,\varepsilon^2}
+\Big (\frac{212}{27}+4\,\zeta(2)\Big )\,\frac{1}{\varepsilon}
-\frac{3182}{81} 
\nonumber\\[2ex]
&&-\frac{10}{3}\,\zeta(2)-\frac{148}{9}\,\zeta(3)\Bigg \}
+n_f\,T_f\,C_F\,\Bigg \{
-\frac{142}{3}+16\,\zeta(3) \Bigg \}\Bigg ]\,,
\end{eqnarray}
with the non-pole constants $f_{g,k}^{(0)}$
\begin{eqnarray}
\label{eqn4.20}
f_{g,1}^{(0)}&=&C_A\,\Big (4+\zeta(2)\Big )\,,
\nonumber\\[2ex]
f_{g,2}^{(0)}&=&n_f\,T_f\,C_A\,\Big (-\frac{3182}{81}-\frac{10}{3}\,\zeta(2)
-\frac{148}{9}\,\zeta(3)\Big ) +n_f\,T_f\,C_F\,\Big(-\frac{142}{3}
\nonumber\\[2ex]
&&+16\,\zeta(3)\Big )
+C_A^2\,\Big (\frac{7723}{81}+\frac{67}{6}\,\zeta(2)
+\frac{11}{9}\,\zeta(3)\Big )\,.
\end{eqnarray}
Notice that there is $C_F \leftrightarrow C_A$ symmetry between
the electromagnetic (quark) and the (pseudo-) scalar Higgs (gluon) vertex
correction for the $\gamma_{K,ii}^{(k)}$ and $f_{i,k}^{(1)}$. From the general
form of the form factor \cite{sen} and \cite{col} one can predict the 
coefficients of the $1/\varepsilon^4$, $1/\varepsilon^3$
and $1/\varepsilon^2$ terms in Eq. (\ref{eqn4.10}). These can be inferred
from the anomalous dimensions $\gamma_{K,ii}$  Eq. (\ref{eqn4.2}) and $\gamma_{ii}$ 
Eq. (\ref{eqn4.3}). However, and this is new, the coefficient of the single
pole term $1/\varepsilon$ can now also be predicted. In the renormalized vertex 
correction it amounts to $1/2\,\gamma_{ii}^{(1)}+f_{i,2}^{(1)}$.
The form for the electromagnetic vertex function was also
studied in \cite{cat} and \cite{stty}. The form for the quark in \cite{cat} 
agrees with ours up to the double pole term. 
The same is true for the form in \cite{stty}, which is exactly equal
to ours because it is derived from \cite{col}. However the Higgs vertex functions
in Eqs. (4.17) and (4.19) were not discussed. Also the structure of the single pole
term of the scattering amplitude was not given in \cite{cat} and  \cite{stty}.
The latter can be found for instance in \cite{bagl} and \cite{bedi} but the relation
between the quark and the gluon was not observed for the single pole term.       
Using our approach for Eq. (3.9) in \cite{bagl} and for Eq. (A.29) in \cite{bedi}
we can express their $H_i^{(2)}$-functions in the following way
\begin{eqnarray}
\label{eqn4.21}
H_i^{(2)}&=&-\frac{1}{8}\,\gamma_{ii}^{(1)}-\frac{1}{4}\,f_{i,2}^{(1)}+\frac{1}{4}\,
\gamma_{ii}^{(0)}\,K+\frac{3}{8}\,C_i\,\beta_0\,\zeta(2)\,,
\nonumber\\[2ex]
i&=&q,g\,, \qquad C_q=C_F \,,\qquad C_g=C_A \,.
\end{eqnarray}
See also Eqs. (3.10), (3.11) in \cite {bagl} and Eqs. (A.30), (A.33) in \cite{bedi}.
It is clear that the structure of the single pole term is now solved. With this
knowledge and the three-loop anomalous dimension \cite{move} we can at least predict the
three-loop vertex function and three-loop scattering amplitudes up to the double pole 
term.

Summarising the above we have calculated the two-loop vertex functions
for the scalar and pseudo-scalar Higgs boson including finite terms. 
We found agreement with \cite{harland}
for scalar Higgs production for $C_A=3$ and $C_F=4/3$ but the pseudo-scalar
Higgs vertex correction is published for the first time. Further we 
confirmed the predictions for the form factor 
up to second order double pole term made in \cite{sen} and \cite{col}.
However we could also extend them to the coefficient of the second order single pole 
term which equals $1/2\,\gamma_{ii}^{(1)}+f_{i,2}^{(1)}$ for the quark ($i=q$) 
and the gluon ($i=g$).\\[3mm]

Acknowledgement: J. Smith would like to thank George Sterman for a discussion of the
above results.

%

\end{document}